\documentstyle[amssymb,prb,aps,epsfig]{revtex}

\begin{document}

\twocolumn[
\hsize\textwidth\columnwidth\hsize\csname@twocolumnfalse\endcsname 

\title{Mott-Hubbard transition in infinite dimensions}
\author{Ning-Hua Tong$^{1},$ Shun-Qing Shen$^{2,1}$ and Fu-Cho Pu$^{1,3}$}
\address{$^{1}$State Key Laboratory of Magnetism, Institute of Physics\\
Chinese Academy of Sciences. P.O. Box 603-12, Beijing 100080, China\\
$^{2}$Department of Physics, The University of Hong \ Kong, Pokfulam Road,\\
Hong Kong, China\\
$^{3}$Department of Physics, Guangzhou Normal College, Guangzhou 510400,\\
China}
\date{\today}
\maketitle

\begin{abstract}
We analyze the unanalytical structure of metal-insulator transition (MIT) in
infinite dimensions. A multiple-valued structure in Green's function and
other thermodynamical quantities with respect to the interaction strength $U$
are observed at low temperatures by introducing a transformation into the
dynamical mean-field equation of Hubbard model,. A complete description of
stable, metastable and unstable phases is established in the regime $U_{c1}(T)<U<U_{c2}(T)$. The Maxwell construction is performed to evaluate
the MIT line $U^{\ast }(T)$. We show how the first-order MIT at $U^{\ast }(T)
$ for $T>0$ evolves into the second-order one at $U_{c2}(0)$ for $T=0$. The
phase diagram near MIT is presented.


\vspace{1cm}
\end{abstract}
]

\section{Introduction}

The Mott-Hubbard metal-insulator transition (MIT) is one of the classic
topics in strongly-correlated electron systems.$^{1,2,3}$ The one-band
Hubbard model is thought to be the minimum one such that the main features
of the MIT are grasped. Early studies by Hubbard, and Brinkman and Rice
explain very well the high and low energy behaviors of the local
single-particle spectrum, respectively.$^{4}$ In recent years, the
transition between a Fermi-liquid phase and an insulating phase is
extensively studied in the framework of dynamical mean-field theory (DMFT),$%
^{5,6}$ by utilizing many methods such as iterative perturbation theory
(IPT),$^{7}$ quantum Monte Carlo (QMC) calculation,$^{8-10}$ exact
diagonalization (ED),$^{11,12}$ projective self-consistent technique (PSCT),$%
^{13}$ and numerical renormalization group method.$^{14,15}$ Below a
critical temperature $T_{c}$,$^{8}$ a paramagnetic (PM) metallic phase and a
PM insulating phase$^{16}$ coexist in the regime $U_{c1}(T)<U$ $<U_{c2}(T)$,$%
^{10,15,17}$ where $U$ is the on-site Coulomb repulsion, and $U_{c1}(T)$ and 
$U_{c2}(T)$ are boundaries of the coexistence regime at temperature $T$.
When $U$ increases, the metallic phase characterized by a finite density of
spectrum at the Fermi-level disappears discontinuously at $U_{c2}(T)$. On
the other hand, the insulating phase in the large $U$ regime is destroyed
abruptly when $U$ decreases to $U_{c1}(T)$.$^{15}$ Between $U_{c1}(T)$ and $%
U_{c2}(T)$ lies a first-order MIT line $U^{\ast }(T)$, at which the free
energies of the two coexisting phases are equal.$^{10,17}$ At the critical
temperature $T_{c}$, both $U_{c1}(T)$ and $U_{c2}(T)$ equal to $U_{c}$ and a
second-order transition occurs. Above $T_{c}$, when $U$ increases, the
system changes from metallic into insulating phase through crossover.$^{17%
\text{,}18}$ At zero temperature, it is found that the metallic phase has a
lower energy in the coexisting regime. The MIT occurs at $U=U_{c2}(0)$ in
such a way that the energy difference between the two coexisting solutions
disappears quadratically, i.e., $E_{g}^{I}-E_{g}^{M}$ $\backsim $ $%
(U_{c2}(0)-U)^{2}$.$^{13}$

Despite the enormous efforts on this problem, there is no complete
description for the stable, metastable and unstable phases. These phases
always appear near the first-order phase transition and are important for
understanding the behaviors of the system under external influences.
Traditionally, the first-order nature of MIT is disclosed from the
hysteresis or discontinuity in some relevant physical quantities with
respect to $U$. In this way, the unanalytical feature of these quantities is
not displayed, and the Maxwell construction cannot be performed explicitly
to produce the MIT line $U^{\ast }(T)$. Hence the MIT deserves further study
in this direction. In this paper, a transformation is introduced to the DMFT
self-consistent equation to investigate the unanalytical structure in the
Green's function, local density of states (DOS) and double occupation
probability in terms of $U.$ \ For each $U$ in the regime $U_{c1}(T)<U$ $%
<U_{c2}(T)$, three solutions arise: stable, metastable and unstable
non-Fermi-liquid metallic solutions. The Maxwell construction is performed
to produce the MIT line $U^{\ast }(T)$. The second-order MIT at $T=T_{c}$ is
found to persist at a metastable level in the low temperature limit. When
temperature approaches zero, the first-order transition at $U^{\ast }(T)$
evolves into second-order one$^{19}$ at $U_{c2}(0)$. Finally an ED phase
diagram of MIT is presented.

\section{Model and method}

We start with the single-band Hubbard model at half-filling for the PM
solution:

\begin{equation}
H=-t\sum\limits_{<i,j>\sigma }c_{i\sigma }^{\dagger }c_{j\sigma
}+\sum\limits_{i}Un_{i\uparrow }n_{i\downarrow }.  \eqnum{1}  \label{1}
\end{equation}
The notations are conventional. A semicircular bare density of states is
used: $D(\varepsilon )=(2/\pi W^{2})\sqrt{W^{2}-\varepsilon ^{2}}$, and $%
W=1.0$ is taken as the energy unit. To introduce our method, denote a
thermodynamical quantity by $Q$. For a specific temperature, the formal
dependence of $Q$ on the on-site interaction $U$ is written as $Q=f_{Q}(U)$.
We consider the analogy between MIT and conventional liquid-gas transition,
which was put forward by Castellani {\it et al}..$^{20}$ In this analogy, $U$
corresponds to the pressure $P$ as the driving force of phase transition,
and $D=\left\langle n_{\uparrow }n_{\downarrow }\right\rangle $ corresponds
to the inverse density $v$. This analogy is helpful for us to understand the
characters of MIT.$^{8}$ In the mean-field treatment of MIT, unanalyticity
of thermodynamical quantities versus $U$ should appear in a similar way as
that appears in the isotherm $v=v(P)$ of van der Waals equation. In this
way, we realize that for an appropriate quantity $Q$, the function $f_{Q}(U)$
should be continuous, but may have $``Z"$- or $``S"$-shaped structure. Now
we focus on the Green's function at an imaginary time $\tau $. As an
example, we choose $Q=G(\beta /2)$. A discontinuous jump in $G(\beta /2)$
was observed when we directly swept along $U$ axis. In order to avoid such
discontinuity, instead of directly calculating $G(\beta /2)=f_{G(\beta
/2)}(U)$ for each $U$, we try to find a solution of $G(\beta /2)$ for the
transformed self-consistent equation,

\begin{equation}
G(\beta /2)=f_{G(\beta /2)}(U-\lambda (A-G(\beta /2))),  \eqnum{2}
\end{equation}
where the parameters $A$ and $\lambda $ are assigned such that $G(\beta /2)$
is single-valued with respect to $U$ even if the function $f_{G(\beta
/2)}(U) $ has a $``Z"$- or $``S"$-shaped structure. Existence of such an
unanalytical structure is strongly indicated by the discontinuity of $%
G(\beta /2)$. After the new equation is solved, the original functional
dependence $G(\beta /2)=f_{G(\beta /2)}(U^{\prime })$ is recovered by
plotting the self-consistent solution $G(\beta /2)$ versus the argument $%
U^{\prime }=U-\lambda (A-G(\beta /2))$. In practice, Eq. (2) is combined
with the DMFT self-consistent equations by iterative calculation. For fixed $%
U$, $A$ and $\lambda $, in each iteration, we first calculate $U^{\prime
}=U-\lambda (A-G(\beta /2))$, where $G(\beta /2)$ is taken from previous
iteration. Take the chemical potential $\mu =U^{\prime }/2$ to realize the
constraint of half-filling. Then $U^{\prime }$ is used to replace $U$ in the
conventional algorithm$^{6}$ to proceed with the calculation in this
iteration (or from initialization). At the end, besides the new set of
parameters for the effective single-impurity Anderson model, a new $G(\beta
/2)$ is also produced. Both of them are used in next iteration. $U^{\prime }$
varies with iteration until the Green's function converges. Then both Eq.
(2) and the DMFT equations will be fulfilled. No additional computational
effort is needed in implementing our scheme. After the convergence is
reached, other quantities, such as $G(\tau )$ for a general $\tau $, the
double occupancy $\left\langle n_{\uparrow }n_{\downarrow }\right\rangle $,
and the local DOS at the Fermi surface $\rho (0)$, can be calculated from
the converged Green's function $G(i\omega _{n})$ and $U^{\prime }$. The
whole functional dependence $Q=f_{Q}(U^{\prime })$ can be obtained by
sweeping $U$ in Eq. (2). This modified self-consistency scheme was used in
the study of phase separation in double-exchange systems.$^{21}$ This method
is also effective to reveal the unanalyticity in MIT, and the final results
are essentially independent of the selection of $A$ and $\lambda $ if only $%
\left| \lambda \right| $ is large enough to ``stretch'' the curve. In this
paper, the finite-temperature ED technique is used to implement the above
scheme.$^{22}$ This technique, originally proposed by Caffarel and Krauth,$%
^{11}$ uses $\chi ^{2}$ fit to obtain new parameters of the effective
Anderson impurity model in the iteration process.

\section{Results and discussion}

The minus imaginary-time Green's function at $\tau =\beta /2$ is plotted
versus $U$ in Fig.1(a) for $T=0.01$, which is lower than the critical
temperature $T_{c}$. It is shown that $-G(\beta /2)=-f_{G(\beta /2)}(U)$ is
indeed continuous with a $``Z"$-shaped structure. Three solutions of $%
G(\beta /2)$ coexist in an extent of $U$. At $T=0.01$, the results evaluated
by ED technique of five and six sites are consistent very well. We find that
our results are independent of sufficiently large $\lambda $, the value of $%
A $, or the initial seeding of iteration. We plot $G(\tau )$ versus $G(\beta
/2)$ as shown in the inset of Fig.1(a). The analytical behaviors of these
curves appear clearly that for each $\tau \in \lbrack 0,\beta ]$, $G(\tau )$
is unanalytical in terms of $U$. Moreover, the coexistence boundaries of $%
G(\tau )$ have the same values irrespective of $\tau $. The Fourier
transformation $G(i\omega _{n})$ as well as all other quantities will have
the same boundary of coexistence, if only they depend on the Green's
functions analytically. So, the boundaries of coexistence $U_{c1}(T)$ and $%
U_{c2}(T)$ are well defined in MIT.

\begin{figure}
\centerline{
\epsfig{figure=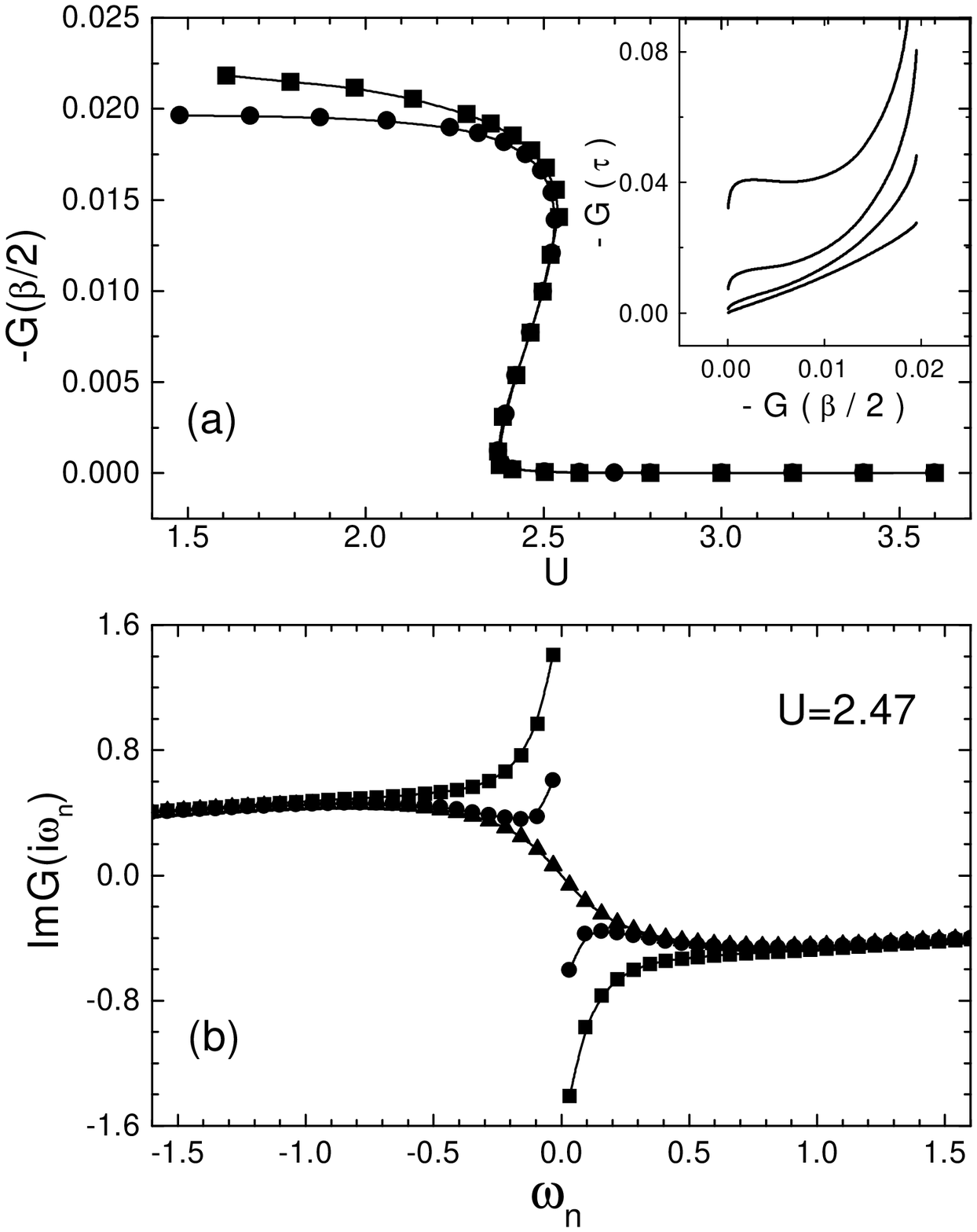,width=9cm}
}
\vspace{-1.5cm}
\caption{
(a)\ $U$-dependence of minus imaginary-time Green's function 
$-G(\tau )$ at $\tau =\beta /2$, obtained using $Ns=5$ (square)and 
$Ns=6$ (dot). (b) The three coexisting solutions of Green's function
on Matsubara frequency axis, obtained using $Ns=6$ at $U=2.47$. 
Both figures are for $T=0.01$, and the thin lines are for guiding 
eyes. Inset of (a): $-G(\tau )$ versus $-G(\beta /2)$ as $U$ varies 
($Ns=6$). From top to bottom, $\tau =\beta/32$, $\beta/16$, 
$\beta/8$, $\beta/4$.
}
\end{figure}

In Fig.1(b), three coexisting solutions of the Matsubara Green's function at 
$T=0.01$ and $U=2.47$\ are presented. The upper and the lower curves are
similar with those from QMC.$^{8,10}$ One is metallic-like (squares) and the
other is insulating-like (up triangles). A new metallic-like solution (dots)
is found between them. At finite temperatures, this newly discovered
solution has the highest free energy among the threes and is unstable. In
this paper, we do not intend to quest for the highest numerical precision,
but focus on the qualitative features of the Mott-Hubbard transition and its
temperature evolution. All our results in the following are obtained using $%
Ns=6$ ($Ns$ is the number of sites).

The minus imaginary-time Green's function $-G(\tau )$ at $\tau =\beta /2$,
the double occupancy $D=\left\langle n_{\uparrow }n_{\downarrow
}\right\rangle $ and the local DOS at Fermi surface $\rho (0)$ are plotted
versus $U$ for several temperatures in Fig.2(a), (b) and (c), respectively.
In Fig.2(a), $-G(\beta /2)$ decreases monotonously when $U$ increases at $%
T=0.04$. As the temperature decreases, it decays more rapidly in the
intermediate $U$ regime. At $T=0.025$, a singular point arises at $U=2.34$
where the slope diverges. Below this temperature, the curve is still
continuous, but has a ``$Z$''-shaped structure. With decreasing temperature,
the curve is compressed along $-G(\beta /2)$ axis and the coexisting regime
extends along the $U$ axis. This leads to a more pronounced multiple-valued
structure. We do not find slowing down of this tendency as temperature is
lowered down to $T=0.0025$. Similar temperature evolution behaviors are also
observed in $D-U$ (Fig.2(b)) and $\rho (0)-U$ (Fig.2(c)) curves. Within our
numerical precision, the three kinds of curves produce\ the same value of
critical point ($U_{c}$, $T_{c}$) and the same boundaries of coexistence
regimes for $T<T_{c}$. This is consistent with our conclusion that the
unanalytical structure is universal for all thermodynamical quantities. The
critical temperature $T_{c}$ and interaction $U_{c}$ are thus estimated to
be $T_{c}\approx 0.025,$ $U_{c}\approx 2.34$, which agree quite well with
that obtained from QMC: $T_{c}=0.026\pm 0.003,$ $U_{c}=2.38\pm 0.02$.$^{8}$
At the critical point, all three quantities have divergent slopes with
respective to $U$. From Fig.2(a)-(c), the boundaries of coexistence regime $%
U_{c1}(T)$ and $U_{c2}(T)$ are easily obtained, in contrast to previous
approaches.$^{10,15,17}$

The double occupancy $D$ is of special thermodynamical significance. The
free energy can be evaluated by integrating along the $D-U$ lines in
Fig.2(b): 
\begin{equation}
F(U,T)=F(0,T)+\int_{0}^{U}D(U^{\prime },T)dU^{\prime }.  \eqnum{3}
\end{equation}
In Fig.2(b), $D-U$ curves have similar unanalytical behavior with $v-P$
isotherms of van der Waals equation. At $T\geqslant T_{c}$, the curves of $%
D-U$ are similar with those from IPT$^{23}$ and QMC.$^{8}$ The Mott critical
point $(U_{c},$ $T_{c})$ was studied in detail by QMC$^{8}$ and Landau
theory of phase transition.$^{23}$ At $0<T<Tc$, three solutions coexist in a
regime around $U_{c}$. We compare free energies of the three solutions for a
fixed $U$ and find that the phase with intermediate $D$ has the highest free
energy. For the metallic (with largest $D$) and the insulating (with
smallest $D$) phases, their free energies cross at the point $U=U^{\ast }(T)$%
, i.e., $F_{M}(U^{\ast },T)=F_{I}(U^{\ast },T)$. We solved this equation
numerically to determine $U^{\ast }(T)$. As $U$ passes by$\ U^{\ast }$ from
below, a stable metallic phase ($F_{M}(U,T)<F_{I}(U,T)$) transits into a
stable insulating phase ($F_{I}(U,T)<F_{M}(U,T)$) (See Fig.3). This
transition is accompanied with a finite jump $\Delta D$ of the double
occupancy $D$, which is determined by the Maxwell construction line as shown
in Fig.2(b). Hence the MIT at finite temperatures is identified as a generic
first-order phase transition. At $T<T_{c}$, it is interesting that there is
a discontinuous jump in the slope of the $D-U$ curve at $U=U_{c1}(T)$. At
this point, $\partial F/\partial U=D$ is continuous but $\partial
^{2}F/\partial U^{2}=$ $\partial D/\partial U$ is discontinuous. It means
that the transition is of second order. This singularity is directly evolved
from that of the critical point $(U_{c},T_{c})$ as temperature decreases. It
turns out that the second-order MIT at $T=T_{c}$ does not disappear in the
regime $T<T_{c}$, but persists to the absolute zero temperature at a
metastable level. This feature of MIT is not unique among the first-order
phase transitions. In the double-exchange model for manganites, similar
feature was observed in the isotherm of charge density $n$ versus chemical
potential $\mu $ (See Fig.2 in Ref. 21). There, when the phase separation
(also a typical first-order phase transition) between paramagnetic (PM) and
ferromagnetic (FM) phases appears at $T<T_{c}$, the second-order PM-FM
transition persists at the metastable level down to zero temperature.

Now we discuss the phenomenon of critical slowing down. In a previous QMC
study of MIT,$^{10}$ the critical slowing down in the convergence of
iteration was observed near the boundaries of coexistence regime, and it was
used as the indicator to determine these boundaries. Such critical slowing
down arises as one tries to directly calculate the ``Z''-shaped curve for
the first-order phase transition. In our calculations we did not observe the
critical slowing down of the same kind since what we practically calculated
is a ``stretched'' single-valued function $f_{Q}(U-\lambda (A-G(\beta /2))$,
and the coexistence of solutions is removed by the transformation. In
contrast, we observed a weak slowing down of convergence at the singular
point at $U=U_{c1}(T)$ where the slope of $D-U$ curve is discontinuous. This
slowing down is associated with the metastable second-order phase transition
identified above. Hence it is different from that discussed in Ref. 10.

\begin{figure}
\centerline{
\epsfig{figure=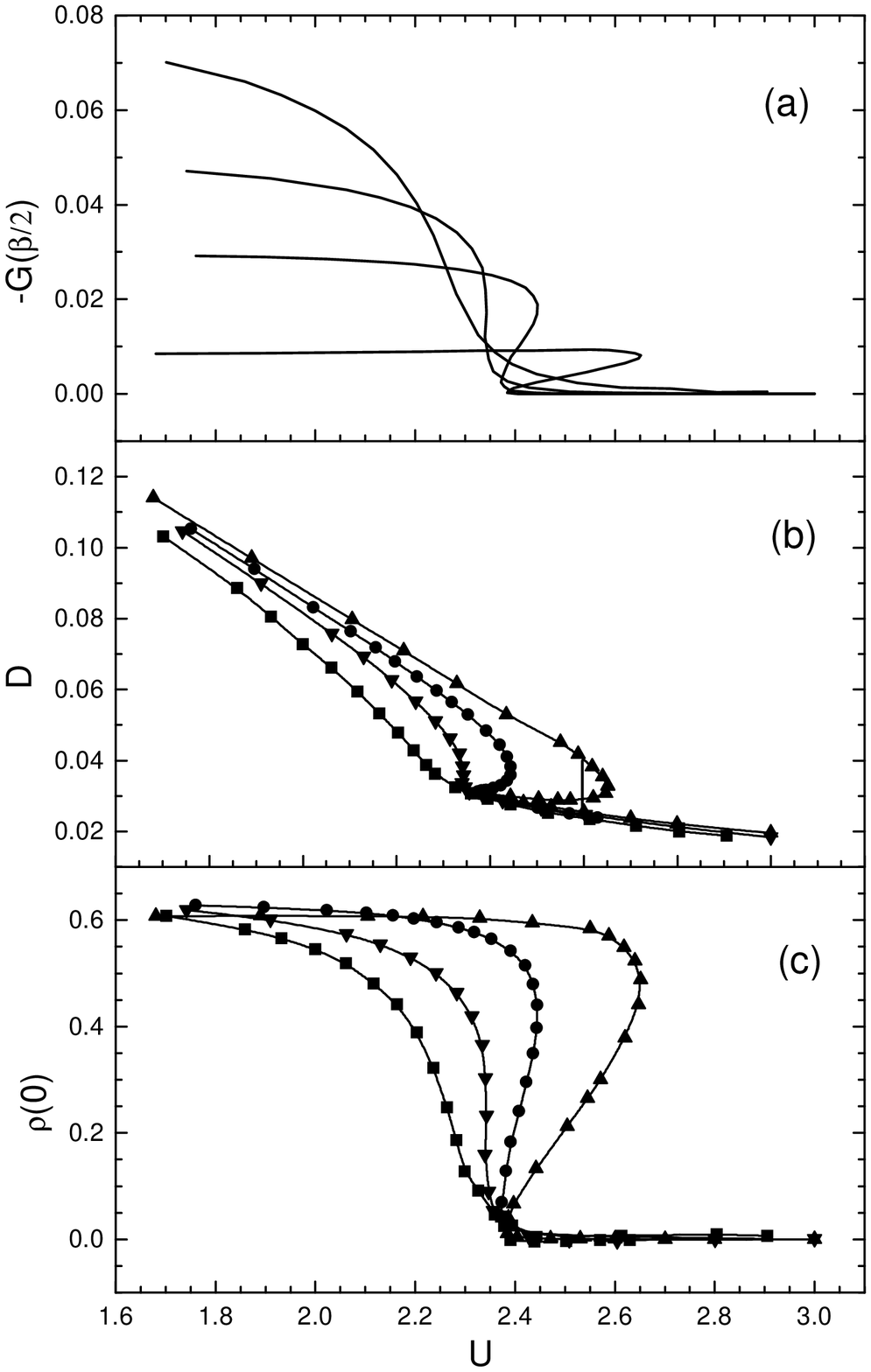,width=10cm}
}
\caption{(a) $U$-dependence of $-G(\beta /2)$ at $T=0.04$, $0.025$,
$0.015$, and $0.005$. (From top to bottom on the left side) (b)
and (c) double occupancy $D$ ( in (b) ) and DOS at Fermi surface 
$\rho (0)$ ( in (c) ) versus $U$ at $T=0.04$ (squares), $0.025$ 
(down triangles), $0.015$ (dots), and $0.005$ (up triangles). 
Thin lines are for guiding eyes. The thick vertical line in (b) 
shows Maxwell construction for $T=0.005$. In small $U$ regime, 
the minor deviation of $\rho (0)$ at $T=0.005$ from $2/\pi W$ originates
from the finite-size effect.
}

\end{figure}

At zero temperature, Rozenberg$^{12}$ and Moeller$^{13}$ found a
second-order MIT at $U_{c}=U_{c2}(0)$. Here, we discuss how the first-order
MIT at finite temperature evolves into the second-order one at zero
temperature. Fig.2(b) shows that when temperature decreases, the linear
behavior of $D-U$ curve becomes dominant and the intermediate branch of $D$
goes closer to the insulating solution. As a result, the first-order
transition point $U^{\ast }(T)$ moves towards $U_{c2}(T)$ according to the
Maxwell construction. This tendency continues when the lowest temperature $%
T=0.0025$ in this paper is reached. It is anticipated that at $T=0$, the
intermediate branch of $D$ merges with the lowest branch and they become
degenerate. The linear metallic $D-U$ line meets the insulating branch at $%
U_{c2}(0)=U^{\ast }(0)$, where the jump of $D$ disappears. Hence the
transition is of second order at $T=0$, and it is consistent with the $D-U$
curves shown in Fig.34 of Ref. 6. The quadratic relation $%
E_{g}^{I}-E_{g}^{M} $ $\backsim $ $(U_{c2}(0)-U)^{2}$ (Ref. 12) is obtained
by assuming a strict linear $D-U$ curve at $T=0$. Here it is worth pointing
out that the term {\it second-order} transition in this context means that
at $U^{\ast }(0)$, $\partial E/\partial U$ is continuous and $\partial
E^{2}/\partial ^{2}U$ is discontinuous, where $E=\min (E_{g}^{I},E_{g}^{M})$%
. This transition is special in that metastable state still exists near the
transition point at zero temperature.

In Fig.2(c), the local DOS at the Fermi surface $\rho (0)$ is calculated by
extrapolating $-ImG(i\omega _{n})/\pi $ to the limit $\omega _{n}\rightarrow
0^{+}$. In the small $U$ regime, $\rho (0)$ does not change with interaction
and remains its non-interacting value $2/(\pi W)$, as required by Luttinger
theorem for momentum independent self-energy.$^{5}$ At temperatures above $%
T_{c}$ (e.g., the curve for $T=0.04$ in Fig.2(c)), the Fermi-liquid phase in
small $U$ regime evolves continuously into an insulating phase in large $U$
regime through a crossover. At temperatures below $T_{c}$, the behavior of $%
\rho (0)$ in the small and large $U$ regime are similar with previous results%
$^{6,15}$. In the coexisting regime, the metallic phase at $U_{c2}$ is
smoothly connected to the insulating phase at $U_{c1}$ through a
non-Fermi-liquid phase. According to the integral along the $D-U$ lines,
this non-Fermi-liquid phase has the highest free energy. In light of the
present result, the discontinuity and hysteresis of $\rho (0)$ in Ref. 15
originates from numerical instabilities at the boundaries of the coexisting
regime. Physically, such instabilities reflect the unstable character of the
metastable phases near phase boundaries. The metastable second-order
transition at $U_{c1}(T)$ is now identified as the transition between
metallic ($\rho (0)>0$) and insulating ($\rho (0)$ $\approx 0$) phases. When
temperature decreases, the intermediate solution of $\rho (0)$ moves
downwards, but there is no obvious tendency that it will merge with the
insulating solution of $\rho (0)$ in the limit $T\rightarrow 0$. This is in
contrast to the $D-U$ curve as shown in Fig.2(b). From Fig.2(c), we conclude
that in the low temperature limit, as $U$ passes by $U_{c2}(0)$ from below, $%
D$ varies continuously but $\rho (0)$ will jump to zero from a finite value
as was observed in NRG study.$^{14}$ Other quantities characterizing the MIT
such as the quasi-particle weight $Z$ (Ref. 11) and local spin-spin
correlation $\left\langle M_{z}(\tau )M_{z}(0)\right\rangle $ (Ref. 17)
should have similar unanalyticity in terms of $U$.

Based on the analysis of the Green's function and other physical quantities,
we can determine the phase diagram near the MIT. In Fig.3, two boundaries of
the coexistence regime meet at a critical point $U_{c}\approx 2.34,$ $%
T_{c}\approx 0.025$. Above this point, the metallic phase in small $U$
regime evolves into an insulating phase in large $U$ regime through a
crossover area. Following Ref. 15, the $U$ regime where $-d\rho (0)/dU$ is
larger than its half maximum is regarded as the crossover regime. Below $%
T_{c}$, between the two boundary lines three phases coexist, i.e., stable,
metastable and unstable phases. Positions and properties of these phases are
labeled in Fig. 3. The MIT line $U^{\ast }(T)$ from Maxwell construction
resembles that obtained by J. Joo {\it et al}..$^{10}$ Compared with the
phase diagram of other authors, the $U_{c2}(T)$ line agrees pretty well with
NRG results$^{15}$, and the $U_{c1}(T)$ line lies between that from NRG$%
^{15} $ and QMC$^{10}$ studies. Although the lowest temperature in the
present paper is $T=0.0025$, our phase boundaries extrapolated to zero
temperature is in good consistency with the results of NRG (Ref.14) and PSCT.%
$^{13}$

\begin{figure}
\centerline{
\epsfig{figure=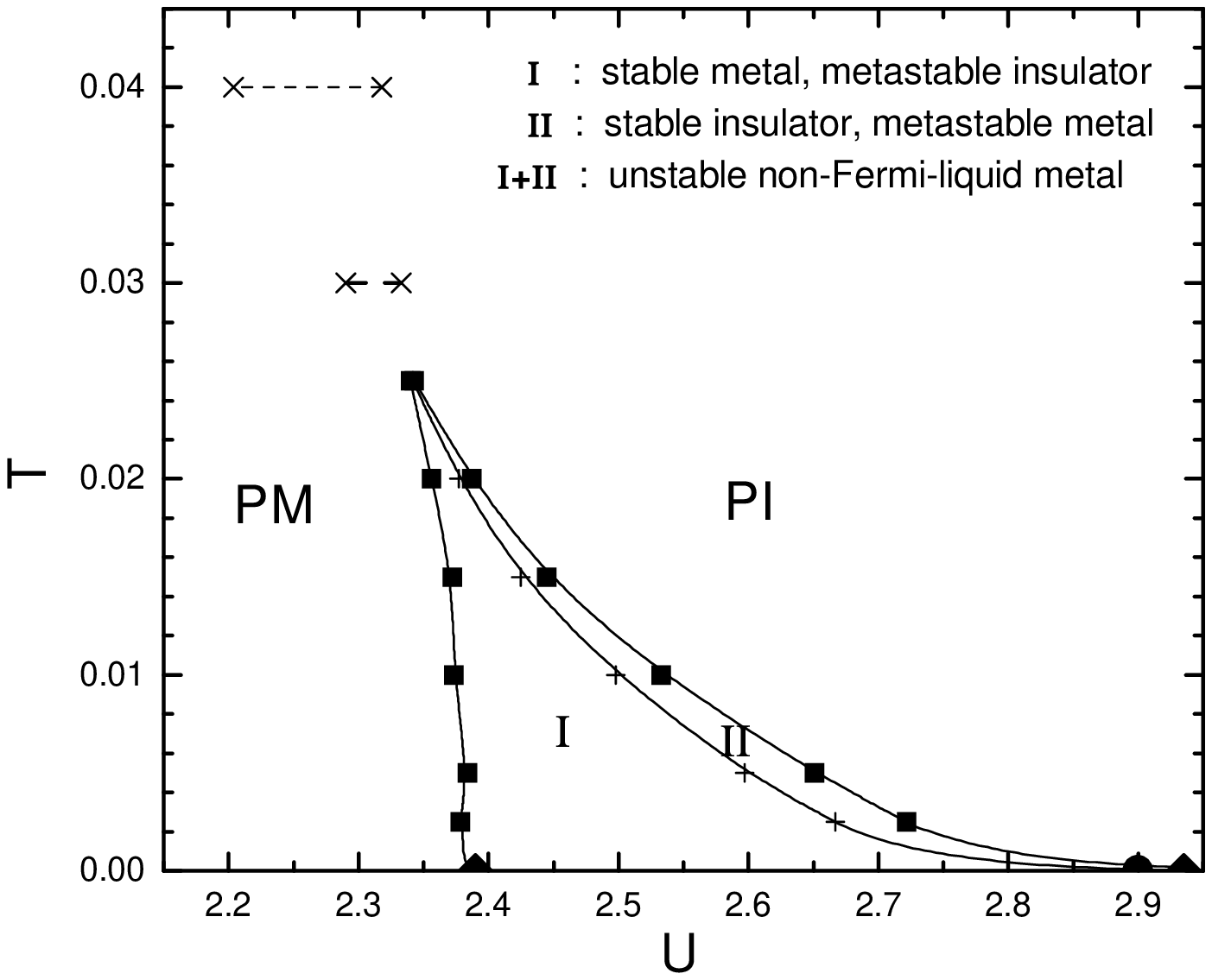,width=14cm}
}
\vspace{-2.5cm}
\caption{
Phase diagram of Mott-Hubbard transition obtained from ED method. 
The squares and crosses mark out the coexistence boundaries and the 
MIT line $U^{\ast }(T)$, respectively. Thin lines are eye-guiding 
lines. Corsses with horizontal dashed lines above $T_{c}$ show the
crossover regime. PSCT$^{(9)}$ (circle) and NRG$^{(10)}$ (diamonds)
results for $T=0$ are also shown. PM, PI denote paramagnetic metal
and paramagnetic insulator, respectively. The properties of phases
in area {\bf {I}}, {\bf {II}} and {\bf {I}}+{\bf {II}} are noted 
in the figure.
}
\end{figure}

\section{Summary}

In summary, we introduce a transformation into the DMFT self-consistent
equations to study the unanalytical behavior of thermodynamical quantities
in the Mott-Hubbard metal-insulator transition in infinite dimensions. Using
the ED technique at finite temperatures, the ``Z''-shaped multiple-valued
structure of several quantities in terms of $U$ is obtained. An unstable
non-Fermi-liquid phase as well as the two phases discovered previously are
found in the regime $U_{c1}(T)<U$ $<U_{c2}(T)$. The MIT line $U^{\ast }(T)$
is obtained by Maxwell construction. The second-order MIT at $T=T_{c}$
persists at the metastable level down to zero temperature, and the
first-order MIT at $U^{\ast }(T)$ evolves into second-order one at $%
U_{c2}(0) $ when temperature approaches zero. The method used in this paper
should be useful for the study of other first-order phase transitions.

\section{Acknowledgment}

This work was supported by a CRCG of the University of Hong Kong and a RGC
grant of Hong Kong.

*Electronic address:

tongnh@aphy.iphy.ac.cn

sshen@hkucc.hku.hk


\begin{references}
\bibitem{16}  Although an insulator is only strictly defined for $T=0$, we
use the term insulating or insulating-like to denote the solution with
vanishingly small spectral weight at the Fermi level for $T>0$.
\bibitem{19}  As to be discussed in the following, the MIT at $T=0$ is of
second order only in the sense that at the transition point, $\partial
E/\partial U$ is continuous and $\partial E^{2}/\partial ^{2}U$ is not. It
is special and should be distinct from the conventional second-order
transition.
\bibitem{22}  Part of the programs provided by Ref. 6 were adapted for our
use.
\end{references}
\end{document}